\documentstyle{mn}

\title{On the r-mode spectrum of relativistic stars}

\author[H.~R.~Beyer and K.~D.~Kokkotas]
{ Horst R.~Beyer$^{1}$ and Kostas D.~Kokkotas$^{2,3}$\\
$^{1}$ Universit\"{a}t Stuttgart, Institut A f\"{u}r Mechanik,
 Pfaffenwaldring 9, D-70550 Stuttgart, Germany.\\
$^{2}$ Department of Physics, Aristotle University of Thessaloniki,
       Thessaloniki 54006, Greece.\\
$^{3}$ Max Planck Institute for Gravitational Physics,
       The Albert Einstein Institute, D-14473 Potsdam, Germany. 
}

\date{Accepted 1999  (?).
      Received 1999  (?);
      in original form  1999}
\begin{document}

\maketitle

\begin{abstract}
We present a mathematically rigorous proof that the $r$-mode spectrum 
of relativistic stars to the rotational lowest order has a continuous part.
A rigorous definition of this spectrum is given in terms of the spectrum of a 
continuous linear operator. 
This study verifies earlier results by Kojima (1998) 
about the nature of the $r$-mode spectrum.
\end{abstract}

\begin{keywords}
stellar oscillations, neutron stars, stellar instabilities
\end{keywords}

\section{Introduction}

The recently discovered $r$-mode instability \cite{nanew,fm}
in rotating neutron stars, has significant implications on the
rotational evolution of a newly-born neutron star.
The $r$-modes are unstable due to the Chandrasekhar-Friedman-Schutz 
(CFS) mechanism \cite{chandra,fs}. 
Two independent computations by Andersson, Kokkotas \& Schutz (1998) 
and Lindblom, Owen \& Morsink (1998)  find that the $r$-mode instability 
is responsible for slowing down a rapidly rotating, newly-born
neutron star to rotation rates comparable to that of the
initial period of the Crab pulsar ($\sim$19 ms).
This is achieved by the emission of current-quadrupole gravitational 
waves, which reduce the angular momentum of the star. 
The instability is active for as long as its
growth-time is shorter than the damping-time due to the viscosity of
neutron-star matter.  
The $r$-mode instability explains
why only slowly-rotating pulsars are associated with supernova
remnants. 
The $r$-mode instability does not allow
millisecond pulsars to be formed after an accretion-induced collapse
of a white dwarf \cite{AKS98a}.  
It seems that millisecond pulsars can only be
formed by the accretion-induced spin-up of old, cold, neutron stars.
Additionally, this instability should be active in 
accreting neutron stars
and so can limit their rotation provided that the stars are hotter than
about $2\times10^5$K and also  in the rapidly 
spinning neutron stars in low mass X-ray binaries (LMXB)\cite{AKS98b}. 
Finally, while the initially rapidly rotating star spins down, an
energy equivalent to roughly 1\% 
of a solar mass is radiated in
gravitational waves, 
making the process an interesting source
of detectable gravitational waves \cite{Owen98}.

Oscillations of stars are commonly described by the Lagrangian
displacement vector $\vec \xi$, which describes the displacement of a
given fluid element due to the oscillation. Since $\vec \xi$ is a
vector on the $(\theta,\phi)$ 2-sphere, it can be written as a sum
of spheroidal and toroidal components (or polar and axial components,
in a different terminology).  In a non-rotating star, the usual $f$,
$p$ and $g$ modes of oscillation are purely spheroidal, characterized
by the indices $(l,m)$ of the spherical harmonic function $Y_l^m$. In
a rotating star, modes that reduce to purely spheroidal modes in the
non-rotating limit, also acquire toroidal components.  Conversely,
$r$-modes in a non-rotating star are purely toroidal modes with
vanishing frequency\footnote{
In the relativistic case the picture is similar \cite{TC67}
nevertheless, in this case there exist an additional family
of quasinormal modes, the 
ones called ``{\em spacetime or $w$-modes}''\cite{chandra91b,kokkotas94}.
}.  

In a rotating star, the displacement vector
acquires spheroidal components and the frequency in the rotating
frame, to first order in the rotational frequency $\Omega$ of the
star, becomes
\begin{equation}
\sigma_r = {{2 m \Omega}\over {l (l+1)}} \ ,
 \label{omega_r}
 \end{equation}
for a given ($l,m$) mode. 
 An inertial observer, measures a frequency of 
\begin{equation}
\sigma_i = m \Omega - \sigma_r.
\label{omega_i}
\end{equation}

As mentioned earlier, when  the star is set in slow rotation 
the axial (toroidal) modes are no longer 
degenerate, but instead the  new family of $r$-modes emerges, 
which are horizontal displacements on equipotential surfaces.
In this case the axial (toroidal) perturbation with spherical harmonic indices
$(\ell,m)$ induce polar (spheroidal) perturbations with harmonic indices
$(\ell \pm 1,m)$ and vice versa.

The picture described above is purely Newtonian, the above calculation 
was performed using Newtonian slowly rotating stellar models and the 
power  radiated in gravitational waves was  estimated using 
the quadrapole formula.
The two approximations (Newtonian theory and slow rotation)
that were used give only a qualitative 
picture while the quantitative results would change if 
at least general relativity is used. 
The assumption of slow rotation is a robust  
approximation because the expansion parameter
$\epsilon =\Omega\sqrt{R^3/M}$ is usually very small and  the fastest
spinning known pulsar has  $\epsilon \sim 0.3$.

The perturbation equations for slowly rotating relativistic stars
were derived by Kojima \shortcite{Kojima92}, see also \cite{chandra91a}.
Kojima (1993) calculated also  the effect  of slow rotation
on $f$-modes.
Andersson (1998)  found the
$r$-mode instability using the same set of equations, although his calculations
overestimate the growth rate of the instability. 

An important difference between Newtonian and general relativistic
calculations is the dragging of the inertial frames, which might 
produce significant changes in the frequency spectra.
This is exactly the case that we are studying in this article.
To be more specific, Kojima \shortcite{Kojima98}
suggested that if one calculates
the $r$-mode frequencies using general relativity to lowest order in 
$\Omega$ the spectrum becomes continuous, in contrast to the 
calculations from Newtonian theory where the spectrum is discrete 
and the frequencies are given  by the formula (\ref{omega_r}).
{\em In this article we prove in a mathematically rigorous way that
Kojima's suggestion for the existence of a continuous part within the
spectrum is
true.}

Continuous spectra have been found in many cases in the past in the
study of differentially rotating fluids 
\cite{SV83,Verdaguer83,Balb84a,Balb84b}. 
The continuous spectrum in these cases were again seen
for the $r$-modes together with a wealth of interesting features 
such as: the passage of low-order $r$-modes from the discrete spectrum
into the continuous one as the differential rotation increases; and the   
presence of low order discrete $p$-modes in the middle of the
continuous spectrum in the more rapidly rotating disks \cite{SV83}.
{\em 
The stars under consideration here have no differential rotation 
and the existence of a continuous part of the spectrum is attributed
to the dragging of the inertial frames due to general relativity.}

\section{Perturbation Equations}

Since our calculations will be based on the equations of Kojima 
\cite{Kojima98,Kojima97}  
which are presented in detail there, 
here we are going only briefly to describe the perturbation equations.

We assume that the star is uniformly rotating with angular velocity
$\Omega\sim O(\epsilon)$ where $\epsilon$, as stated earlier,
is small compared to unity. 
The metric is given by: 
\begin{eqnarray}
ds^2=&-&e^{\nu} dt^2 + e^{\lambda} d r^2 + r^2 (d\theta^2 + \sin^2 \theta
d \phi^2) \cr
&-&2\omega r^2 \sin^2 \theta d t d \phi \ ,
\end{eqnarray}
where $\omega \sim O(\epsilon)$ describes the dragging of the inertial frame.
If we include the  effects of rotation only to order $\epsilon$ the 
configuration is still spherical, because the deformation is of 
order $\epsilon^2$  \cite{Hartle}.
The star then is described by the standard Tolman-Oppenheimer-Volkov (TOV)
equations (cf Chapter 23.5 \cite{TOV}) plus an equation for $\omega$
\begin{equation}
\left(jr^2\varpi'\right)' -16\pi(\rho+p)e^\lambda j r^4 \varpi =0 \ ,
\end{equation}
where we have defined
\begin{equation}
\varpi = \Omega-\omega
\end{equation}
a prime denotes derivative with respect to $r$, and
\begin{equation}
j=e^{-(\lambda+\nu)/2}.
\end{equation}
In the vacuum outside the star $\varpi$ can be written
\begin{equation}
\varpi=\Omega -{{2 J} \over r^3} \ ,
\end{equation}
where $J$ is the angular momentum of the star.
The function $\varpi$, both inside and outside the star is a function of $r$ 
only and continuity of $\varpi$ at the boundary (surface of the star, $r=R$) 
requires that $\varpi_R'=6JR^{-4}$. 
Additionally, $\varpi$ is monotonically increasing function of $r$
limited to
\begin{equation}
\varpi_0 \leq \varpi \leq \Omega ,
\end{equation}
where $\varpi_0$ is the value at the center.

For the study of the perturbations of slowly rotating relativistic stars
we should expand all perturbation functions in spherical harmonics and 
additionally assume a harmonic dependence of time 
$\exp[-i(\sigma t- m\phi)]$. 
Here, $\sigma$ is,
the normalized in units $(M/R^3)^{1/2}$, oscillation frequency.
To lowest order for rotating stars, in accordance with the Newtonian theory,
we expect that the toroidal perturbations
of the fluid have finite frequencies of order $\Omega$ (or $\epsilon$).
Then from the six functions $h_0, h_1, H_0, H_1, H_2, K$ describing the
metric perturbations in the Regge-Wheeler gauge \cite{TC67}, 
only one,  $h_0$, is of the same order as the
function describing the toroidal fluid motions $U$
\footnote{The toroidal displacement vector is defined as
$$
\vec \xi =\left(0, U_{\ell m} \sin^{-1}\theta \partial_{\phi},
             -U_{\ell m} \partial_{\theta} \right) Y^m_{\ell}
$$
}.
All other perturbation functions
plus the variations of the pressure $\delta p$ and density $\delta \rho$
are of higher order and thus  can be omitted. 
We should point out that this approximation is valid only for the
study of the $r$-modes and is not appropriate for the study of other fluid
or  $w$-modes. 

In a recent article Lockitch \& Friedman \shortcite{LF98}
suggest that rotation mixes in general all axial and polar 
perturbations and introduce the
idea of {\em hybrid} modes. 
This mixing also eliminates all purely polar modes, but there is still a set of 
purely axial modes. 
They suggest that this set of purely axial modes should not
exist for relativistic slowly rotating stars at least in the barotropic case. 
As we show here in the specific approximation described above
it is possible that the $r$-mode spectrum has a continuous part.

Using the above assumptions the master equation governing quasi-toroidal
oscillations is given by \cite{Kojima98,Kojima97}
\begin{equation} \label{kojimasequation}
 q \Phi +(\varpi-\mu)\left[v \Phi - \frac{1}{r^4j} 
\left(r^4j \Phi ^{\prime} \right)^{\prime}\right] = 0 \ ,
\end{equation} 
where 
\begin{equation}
\Phi = {h_0 \over r^2} \ ,
\end{equation}
and 
\begin{equation} \label{v}
v= \frac{e^\lambda}{r^2} (l-1)(l+2) \ ,
\end{equation}
\begin{equation} \label{q}
q= \frac{1}{r^4j} \left(r^4j \varpi^{\prime} \right)^{\prime}
  = 16\pi(\rho+p)e^\lambda \varpi \ ,
\end{equation}
\begin{equation} \label{mu}
\mu = -\frac{l(l+1)}{2m}(\sigma-m\Omega).
\end{equation}
Equation (\ref{kojimasequation}) in the Newtonian limit 
($j\rightarrow 1$, $q \rightarrow 0$ and $\varpi \rightarrow \Omega$) 
reduces to a simple condition for the $r$-mode frequency 
$\varpi - \mu =0$ which is identical with (\ref{omega_i}).
To this order of $\Omega$ the eigenfrequency of the $r$-modes is
independent of the radial dependence of the eigenfunction $\Phi(r)$
(or $h_0(r)$) and in this sense the eigenfrequency is infinitely
degenerate.

The master equation (\ref{kojimasequation}) is not a regular eigenvalue
problem since the coefficient $(\varpi-\mu)$ becomes singular inside the star
for a certain value of $\mu$. 
{\em The purpose of this work is to study the spectrum of this equation in 
a rigorous way.}

\subsection{Conventions}
The following {\bf conventions} are used: \\ 
The symbols $N$,  $I$, $C$  denote the natural numbers 
(including zero),  all real numbers greater than zero and the 
complex numbers, respectively. 
With $r$ we will denote interchangeably some chosen element 
from $I$ or else the 
identical mapping from $I$ to the real numbers. 
The definition used
will be clear from the context.  For each $k \in N$ the symbol
$C^k(I,C)$ denotes the linear space of $k$-times 
continuously differentiable complex-valued functions on $I$. 

Throughout the paper Lebesgue integration theory 
in the formulation of \cite{RieszNagy} is used.  
Compare with respect to this also Chapter III in 
\cite{HirzebruchScharlau} and Appendix A in \cite{Weidmann}.
Following common usage there is no difference made between an 
almost everywhere  (with respect to the chosen measure) defined 
function $f$ and its associated  equivalence class
(consisting of all almost everywhere defined functions which differ 
from $f$ only on a set of measure zero). 
In this sense
$L_{C}^2\left(I,r^4j\right)$   
denotes the Hilbert space of complex-valued, 
square integrable functions (with respect to the measure $r^4j \, dr$)
on the real line. 
The scalar product 
$<| >$ on $L_{C}^2\left(I,r^4j\right)$  is defined by
\begin{equation} \label{scalarproduct}
<f|g> := \int^{\infty}_{0}
   r^4 j \, f^*
   g dr \,
\end{equation}
for all $f,g \in L_{C}^2\left(I,r^4j\right)$.
Finally, $L_{C}^2\left(I^2\right)$ 
denotes the Hilbert space of complex-valued, with respect to the Lebesgue measure 
square integrable functions on the two-dimensional interval $I^2$.

\section{The spectrum of Kojima's master equation}

In the following and in the remainder of the paper we consider the 
cases $l \geq 1$ only. 
Then, in a  first step,  we rewrite (\ref{kojimasequation}) by introducing a new 
dependent variable $\varphi$ defined by: 
\begin{equation} \label{varphiofPhi)}
\varphi :=  v \Phi - \frac{1}{r^4j} 
\left(r^4j \Phi ^{\prime} \right)^{\prime} \, .
\end{equation}
The corresponding representation of 
$\Phi$ in terms of $\varphi$ can be performed with the help of special 
linear independent solutions $\Phi_1 , \Phi_2$ of the differential equation
\begin{eqnarray} \label{kojimasequation1}
&& \frac{1}{r^4j} \left(r^4j \Phi ^{\prime} \right)^{\prime} -v \Phi = \cr  
& & \Phi^{\prime\prime} + \left(\frac{4}{r}+\frac{j^{\prime}}{j}\right)
\Phi^{\prime}-\frac{(l-1)(l+2)}{r^2} e^{\lambda} \Phi = 0 \ , 
\end{eqnarray}
given in the next section.
This representation is given by:
\begin{eqnarray} \label{Phiofvarphi}
\Phi(r) \qquad := 
&&-\frac{1}{W}\left[ \Phi_{2}(r)\int_{0}^{r} \Phi_{1}r^{\prime 4}j\varphi
dr^{\prime} \right. \cr 
&& + \left. \Phi_{1}(r)\int_{r}^{\infty} \Phi_{2}r^{\prime 4}j\varphi
dr^{\prime} \right] \, ,
\end{eqnarray}
where $W$ is the Wronskian  of $\Phi_{1}$ and $\Phi_{2}$, defined by
\begin{equation} \label{W}
W := r^4j\left(\Phi_{1}\Phi_{2}^{\prime}-\Phi_{1}^{\prime} \Phi_{2}\right) \, .
\end{equation}
Roughly, $\Phi_1$ is an element of $L_{C}^2\left(I,r^4j\right)$  for small $r$
and $\Phi_2$ is an element of $L_{C}^2\left(I,r^4j\right)$  for large $r$. 
Apart from an irrelevant factor
such functions are uniquely defined by these demands. The use of these
functions in the inversion 
(\ref{Phiofvarphi}) of (\ref{varphiofPhi)}) is necessary because
of the demand (boundary condition) that $\Phi$ is an element of 
$L_{C}^2\left(I,r^4j\right)$. 

Introducing the new variable $\varphi$ into (\ref{kojimasequation}) 
leads to the following  equation for $\varphi$:
\begin{eqnarray} \label{kojimanew}
&&-\frac{q}{W}\left[\Phi_{2}(r)\int_{0}^{r} \Phi_{1}r^{\prime 4}j\varphi
dr^{\prime} 
+\Phi_{1}(r)\int_{r}^{\infty} \Phi_{2}r^{\prime 4}j \varphi
dr^{\prime} \right] \cr 
&&+ (\varpi(r) -\mu) \varphi(r) = 0 \, ,
\end{eqnarray} 
for all $r \in I$. 

In a second step (\ref{kojimanew}) is turned into a regular spectral problem for 
the continuous linear operator $A$ on  $L_{C}^2\left(I,r^4j\right)$ 
defined as follows:
\newline
For every $f \in L_{C}^2\left(I,r^4j\right)$ we define 
$Af  \in L_{C}^2\left(I,r^4j\right)$ by 
\begin{eqnarray} \label{A}
(Af)(r) :=  &-&\frac{q}{W}\left[\Phi_{2}(r)\int_{0}^{r} \Phi_{1}r^{\prime 4}jf
dr^{\prime} \right. \cr
&+& \left. \Phi_{1}(r)\int_{r}^{\infty} \Phi_{2}r^{\prime 4}jf
dr^{\prime} \right]+ \varpi(r)f(r) \, ,
\end{eqnarray} 
for all $r \in I$. 
That this indeed defines a continuous linear operator on 
the whole of $L_{C}^2\left(I,r^4j\right)$ can be seen as follows. 
First, obviously,  since $\varpi$ is bounded continuous, by 
\begin{equation} \label{Tvarpi}
T_{\varpi}f := \varpi f  \, ,  f \in  L_{C}^2\left(I,r^4j\right)
\end{equation}
there is defined a continuous linear operator on $L_{C}^2\left(I,r^4j\right)$. 
Second, in the Appendix A it will be shown that from
\begin{eqnarray} \label{perturbation}
(Bf)(r) :=  &-&\frac{q}{W}\left[\Phi_{2}(r)\int_{0}^{r} \Phi_{1}r^{\prime 4}jf
dr^{\prime} \right. \cr
&+& \left. \Phi_{1}(r)\int_{r}^{\infty} \Phi_{2}r^{\prime 4}jf
dr^{\prime} \right] \, ,
\end{eqnarray}  
for all $r \in I$ and every $f \in  L_{C}^2\left(I,r^4j\right)$
there is even defined a Hilbert-Schmidt operator $B$ on 
$L_{C}^2\left(I,r^4j\right)$.
Hence $B$ is not only continuous but in addition {\it compact} and Hilbert-Schmidt. 
As a consequence (\ref{A}) defines a continuous linear operator on  
$A$ on $L_{C}^2\left(I,r^4j\right)$, being equal to the sum of $T_{\varpi}$
and $B$. 

The determination of the spectrum of (\ref{kojimasequation}) is now reduced
to finding the spectrum $ \sigma(A)$ of the continuous  linear (non-self-adjoint)
operator $A$. 
Since $A$ is continuous it follows that  $ \sigma(A)$ is a 
{\it bounded} subset of the  complex plane contained in a circle around the 
origin with the radius given by the 
operator norm  $\| A \|$ of $A$ \cite{ReedSimon78}.
The so called essential spectrum of the operator 
$T_{\varpi}$ is given by the values
\begin{equation} \label{spec}
\varpi_0 \leq \mu = - {{\ell(\ell+1)}\over {2m}}\left(\sigma -m\Omega\right)
\leq \Omega \, .
\end{equation} 
(see \cite{ReedSimon78,ReedSimon80})
Moreover it is known (see e.g. \cite{ReedSimon78} 
Corollary 2 of Theorem XIII.14 in Vol. IV) 
that the essential spectrum 
\footnote{Defined according to  \cite{ReedSimon78} on  p. 106 of Vol. IV \, . }
is invariant under perturbations by a compact linear operator such as $B$. 
Hence the essential spectrum  $\sigma_{ess}(A)$ of $A$, which is a part 
of $ \sigma(A)$, is also given by the range of values (\ref{spec}). 
The complement  $ \sigma_{disc}(A) := \sigma(A) \setminus  \sigma_{ess}(A)$  
(which is possibly empty) consists of  {\it isolated} eigenvalues of 
finite multiplicity  \cite{ReedSimon78}.
Using an argument of Kojima \shortcite{Kojima98} it follows that these 
eigenvalues are real (hence $ \sigma(A)$ is also real) and are contained in 
the interval  $(\Omega, \| A \|]$.

\section{Determination of $\Phi_1$ and $\Phi_2$}

Here we distinguish two cases. 
\newline
\newline
The first  uses the following  assumptions on $\rho$ and $p$ \, : \\
Both,  $p$ and $\rho$ are continuous real-valued functions on $I$
satisfying 
the Tolman-Oppenheimer-Volkov equations and in addition are such that
the limits
\begin{equation} \label{behavorigin}
 \lim_{r \to 0} \rho(r)  \quad \mbox{\rm and}  \quad  \lim_{r \to 0} p(r) 
\end{equation}
both exist and 
\begin{equation} \label{behavasympt}
 \rho(r) = 0  \quad \mbox{\rm and} \quad  p(r) 
= 0 \quad \mbox{\rm for} \quad r \geq R \, ,
\end{equation}
are both satisfied, where $R$ denotes the radius of the star. 
\footnote{These assumptions 
are satisfied for instance for slowly rotating polytropic stellar models 
(see e.g., \cite{Hartle}).}
Under these conditions 
the existence of the special solutions $\Phi_1 , \Phi_2$ of the 
differential equation
(\ref{kojimasequation1}) is given by a theorem of Dunkel \cite{Dunkel} 
(compare also \cite{Levinson}, \cite{Bellman} and \cite{Hille})
on linear first-order systems of differential equations having 
asymptotic constant coefficients at $+\infty$. 
By transformation one gets 
from this a theorem where the singular point is finite. 
This theorem - which for the
reader's convenience is given in the Appendix A -  generalizes 
well - known results on weakly singular linear first-order systems
with analytic coefficients.
The  determination of $\Phi_1$ and $\Phi_2$ proceeds now as follows.
First from the TOV equations it follows that $\lambda,j$ 
are continuously differentiable  functions on $I$ such that both:
\begin{equation} \label{asymptotics1} 
\frac{j^{\prime}}{j} \quad \mbox{\rm and} \quad 
\frac{e^{\lambda}-1}{r} 
\end{equation}
are  continuous as well as {\it Lebesgue integrable near} $0$ and that both
\begin{equation} \label{asymptotics2} 
r^2 \, \frac{j^{\prime}}{j} \quad \mbox{\rm and} \quad 
r \left(e^{\lambda}-1\right) 
\end{equation}
are continuous as well as {\it Lebesgue integrable near} $\infty$.
As a consequence of the theorem given in the Appendix A follows 
the existence of linearly independent solutions
$\Phi_{1}, \Phi_{3}$ of  (\ref{kojimasequation1}) satisfying 
\begin{eqnarray} \label{specialsol1}
\lim_{r \rightarrow 0}  r^{1-l}\Phi_{1}(r) &=& 1  \cr
\lim_{r \rightarrow 0}  r^{2-l}\Phi_{1}^{\prime}(r) &=& l-1 \cr
\lim_{r \rightarrow 0}  r^{l+2}\Phi_{3}(r) &=& 1  \cr
\lim_{r \rightarrow 0}  r^{l+3}\Phi_{3}^{\prime}(r) &=& -(l+2) 
\end{eqnarray}
and of linearly independent solutions 
$\Phi_{2}, \Phi_{4}$ of  (\ref{kojimasequation1}) satisfying 
\begin{eqnarray} \label{specialsol2}
\lim_{r \rightarrow \infty} r^{l+2}\Phi_{2}(r) &=& 1  \cr
\lim_{r \rightarrow \infty} r^{l+3}\Phi_{2}^{\prime}(r) &=& - (l+2) \cr
\lim_{r \rightarrow \infty} r^{1-l}\Phi_{4}(r) &=& 1  \cr
\lim_{r \rightarrow \infty} r^{2-l}\Phi_{4}^{\prime}(r) &=& l-1 \, .
\end{eqnarray}

In the next step we conclude that $\Phi_{1}, \Phi_{2}$ are linear independent.
The remaining solutions $\Phi_{3}, \Phi_{4}$ will be  important in  the 
proof of the compactness of $B$ (see appendix).  

Up to now everything said in this section is also valid for the case $l=0$. 
For the following proof of linear independence we have to assume that $ l \geq 1 $.
This is assumed in the remainder of the paper.

The proof of the linear independence of $\Phi_{1}, \Phi_{2}$  
proceeds indirectly. 
So we assume on the contrary that there
is a non-vanishing real $\alpha$ such that $\Phi_{1} = \alpha \Phi_{2}$. 
Using this along with
(\ref{specialsol1}), (\ref{specialsol2}),  
Lebesgue's dominated convergence theorem and the monotonous 
convergence theorem we conclude:
\begin{eqnarray} \label{linindep}
0 &=&\int_{0}^{\infty} \Phi_{1}
\left[-\left(r^4 j\Phi_{1}^{\prime}\right)^{\prime}+
(l-1)(l+2)r^2je^{\lambda}\Phi_{1}\right] dr \cr
&=&\lim_{n \rightarrow \infty} \left[-r^4j\Phi_{1} \Phi_{1}^{\prime} 
\left|_{R/n}^{nR} \right. 
+ \int_{R/n}^{nR} r^4j \Phi_{1}^{\prime 2} dr \right] \cr 
&+& (l-1)(l+2) \int_{0}^{\infty}r^2 je^{\lambda}\Phi_{1}^{2} dr \cr
&=& \int_{0}^{\infty}r^4j\Phi_{1}^{\prime 2} dr 
+  (l-1)(l+2) \int_{0}^{\infty}r^2 
je^{\lambda}\Phi_{1}^{2} dr \, . \nonumber
\end{eqnarray}
Where we have made use of the fact that the function 
\begin{equation}
r^2je^{\lambda}\Phi_{1}^{2}
\end{equation}
is Lebesgue integrable on $I$. This can be concluded from 
the facts that the function $j$ has a continuous extension onto 
the closed interval
$[0,\infty)$, that the functions $\lambda$ and $j$ 
are constant for $r \geq R$ and
that 
\begin{eqnarray} \label{asymptotics3}
\lim_{r \rightarrow 0} \, \, e^{\lambda(r)} &=& 1 \cr
\lim_{r \rightarrow \infty}  e^{\lambda(r)} &=& 1 \, .
\end{eqnarray} 
All these facts are consequences of the TOV equations and the assumptions 
made on  $p$ and $\rho$.  
Hence it follows from (\ref{linindep}) the {\bf contradiction} that 
the function $\Phi_{1}$ is trivial. As a consequence the functions 
$\Phi_{1}, \Phi_{2}$ are linear independent.
\newline \newline
The second case considers slowly rotating homogeneous stellar models.
In this case the functions
$e^{\lambda}$ and $j$ can be given explicitly in terms of 
elementary functions (see e.g., \cite{ChandraMiller}). 
It turns out that $\lambda,j$ are continuous functions and that $j$ is
continuously differentiable on $I\setminus \{R\}$  such that the 
limits at $R$ of the derivative of $j$ 
from the right and from the  left, respectively,
differ from each other. The functions (\ref{asymptotics1}) are also 
{\it Lebesgue integrable near} $0$ and (\ref{asymptotics2}) are also 
{\it Lebesgue integrable near} $\infty$. 
In addition (\ref{asymptotics3}) is also valid for this case. 
As a consequence of the theorem in the Appendix A follows the existence of 
continuously differentiable functions 
$\Phi_{1}, \Phi_{2}$ on $I$ which 
are two times continuously differentiable on 
$I\setminus \{R\}$, satisfy
(\ref{kojimasequation1}) on 
$I\setminus \{R\}$ and in addition satisfy 
(\ref{specialsol1}), (\ref{specialsol2}). 
The linear independence of these functions can be proved completely 
analogous to the previously considered case.

\section{Discussion}
In the previous section we showed for a wide class of background models for
slowly rotating stars that for each $r_0 \in I$ the corresponding
$\mu = \varpi(r_0)$ (or rather the corresponding $\sigma$ according to 
(\ref{mu})) belongs to the spectrum of (\ref{kojimasequation}).  
Furthermore we achieved a 
{\it precise definition for the spectrum of} (\ref{kojimasequation}) 
{\it as the spectrum of the linear operator A}.

This is an interesting new result that needs further study. 
There are still questions to be answered in order to 
understand the effect of the frame dragging induced by general relativity
to the spectrum of the rotating stars. 
For example, one cannot exclude the possibility
that isolated eigenvalues might also exist \cite{nils}. 
Also, the specific form of equation (\ref{kojimasequation}) is found under the
assumption that the $r$-mode frequencies are of order  $\Omega$. 
This implies that a few terms involving toroidal motions and all 
the terms related to spheroidal motions have been omitted because 
they were either either of higher order or because they contribute via 
higher harmonics $(l\pm 1,m)$. 
This is not the case with the hybrid modes
found by Lockitch \& Friedman \shortcite{LF98} where the spheroidal
motions have been taken into account. 
If one includes these extra terms the form of the equation will change and 
the effect on the spectrum will be the emergence of the hybrid modes, 
but still the underlying nature of the spectrum can be the same and
the techniques applied here can be used as well.

Moreover, in the more general case, there is an imaginary part for each 
frequency which corresponds to the damping or growth of the fluid motions 
due to the emission of gravitational radiation or due to the viscosity, 
so it will be interesting to examine whether the spectrum will still be
continuous.

In other words the discovery of the existence of a continuous part 
of the $r$-mode spectrum is just the first step towards understanding the 
real nature of the spectrum. 
Already the idea of hybrid modes \cite{LF98,LI98} adds new features
in the spectrum that have been overlooked in all previous studies. 
More work towards identifying possible isolated
eigenvalues of the spectrum should be done and additionally there is need
to examine if this specific nature of the spectrum is preserved when 
terms of order $\Omega^2$ will be included in equation 
(\ref{kojimasequation}), 
the coexistence of continuous and discrete part of the spectrum is 
not at all impossible.

Concluding, we would like to point out that the existence of a continuous part 
of the spectrum might affect some of the astrophysical estimations 
being made for the growth time of the $r$-mode instability 
and consequently all the Newtonian estimations being made earlier
by \cite{AKS98a,KS98,AKS98b}.

\section*{Acknowledgments}
We would like to thank N.~Andersson, J.~L.~Friedman, K.H.~Lockitch, 
B.~G.~Schmidt, B.~F.~Schutz, and N.~Stergioulas for helpful discussions, 
during our visit to Albert Einstein Institute, Potsdam. 
We are grateful to G.~Allen for the critical reading of the manuscript.


\appendix

\section{PROOF OF THE THEOREM}

The variant of the theorem of Dunkel \cite{Dunkel} 
(compare also \cite{Levinson,Bellman,Hille})
used in Section~4 is the following.

{\it Theorem}: Let $n \in N$; $a, t_0 \in R$ with $a < t_0$; $\mu \in N$;
$\alpha_{\mu} := 1$ for  $\mu=0$ and
$\alpha_{\mu} := \mu$ for  $\mu \neq 0$. 
In addition let $A_0$ be a diagonalizable  complex  $n \times n$ 
matrix and $e^{\prime}_{1} , \dots,   e^{\prime}_{n}$
be a basis of $C^{n}$ consisting of eigenvectors of  $A_0$. 
Further, for each $j \in \{1,\cdots,n\}$ let $\lambda_{j}$ 
be the eigenvalue corresponding to  $e^{\prime}_{j}$ and $P_{j}$ 
be the matrix representing the projection of $C^{n}$
onto $C.e^{\prime}_{j}$ with respect to the canonical basis of $C^{n}$.
Finally, let $A_{1}$ be a continuous map from $(a, t_0)$ into the complex 
$n \times n$ matrices $M(n \times n,C)$ for which there is 
a number $c \in (a,t_0)$ such that the restriction 
of $A_{1jk}$ to $[c,t_0)$ is Lebesgue integrable for each 
$j,k \in {1,..., n}$. 
\newline
\newline
Then there is a $C^{1}$ map $R:(a,t_{0}) \rightarrow M(n \times n,C)$
with $lim_{t \rightarrow 0}R_{jk}(t) = 0$ for each $j,k \in {1,\dots,n}$ 
and such that $u:(a,t_{0}) \rightarrow M(n \times n,C)$ defined by
\begin{eqnarray} \label{u}
&u(t):= \cr
&\left\{
 \begin{array}{ll}
  \sum^{n}_{j=1}(t_0-t)^{-\lambda_{j}}\cdot (E+R(t))\cdot P_{j} 
  \ \mbox{for $\mu=0$} \cr
  \sum^{n}_{j=1}exp(\lambda_{j}(t_0-t)^{-\mu})\cdot (E+R(t))\cdot P_{j} 
   \ \mbox{for $\mu \neq 0$ }
 \end{array} \right. \,
\end{eqnarray}
for all $t \in (a,t_{0}) $
(where $E$ is the $n \times n $ unit matrix),  maps into the 
invertible  $n \times n $ matrices and satisfies
\begin{equation} \label{asymptoticofu}
u^{\prime}(t) = \left( \frac{\alpha_{\mu}}{(t_0-t)^{\mu+1}}A_{0} + A_{1}(t) 
\right) \cdot u(t)
\end{equation}
for each $t \in (a,t_{0})$.
\newline
\newline
Now we show that by (\ref{perturbation}) for all  $r \in I$ and 
every $f \in  L_{C}^2\left(I,r^4j\right)$ there is defined a Hilbert-Schmidt 
operator $B$ on $L_{C}^2\left(I,r^4j\right)$. 
Using the unitary transformation $U$ from $L_{C}^2\left(I,r^4j\right)$
to $L_{C}^2\left(I\right)$ given by 
\begin{equation} \label{U}
Uf := r^2 \sqrt j f \ , \qquad f \in  L_{C}^2\left(I,r^4j\right)
\end{equation}
it easily seen that this is equivalent to showing that
the integral operator $Int(K^{\prime})$ with the kernel
function $K^{\prime}$, where
\begin{eqnarray} \label{kernelprime}   
&K^{\prime}(r,r^{\prime}) := \cr
&-\frac{q(r)(r r^{\prime})^2 (j(r)j(r^{\prime}))^{1/2}}{W} \left\{
\begin{array}{ll}
 \Phi_{2}(r)\Phi_{1}(r^{\prime}) \ \mbox{\rm for $r^{\prime} \leq r$} \\
\Phi_{1}(r)\Phi_{2}(r^{\prime}) \ \mbox{\rm for $r^{\prime} > r $} .
 \end{array} \right.
\end{eqnarray}
for all $r \in I$ and $r^{\prime} \in I$,  defines a 
Hilbert-Schmidt operator on $L_{C}^2\left(I\right)$. It 
is well-known (see e.g. \cite{ReedSimon78,ReedSimon80}), 
that this is equivalent to showing
that $K^{\prime}$ is an element of $L_{C}^2\left(I^2\right)$. 
Since $K^{\prime}$ is continuous this follows if we can show that $|K^{\prime}|^2$
is integrable over $I^2$. For this we notice that as a consequence of (\ref{specialsol1}),
(\ref{specialsol2}), there are positive real $c_1, c_2$, such that 
\begin{eqnarray} \label{est} 
|\Phi_{1}| &\leq& c_{1} r^{l-1}  \cr
|\Phi_{2}| &\leq& c_{2} r^{-(l+2)} \, .
\end{eqnarray}
From this and since $j$ is (by the TOV equations) bounded, 
the integrability of $|K^{\prime}|^2$ follows as an application
of Lebesgue's dominated convergence theorem if the integrability of the
following auxiliary function $H$ can be shown:
\begin{equation} \label{H}   
H(r,r^{\prime})  := 
|q(r)|^2\left\{
\begin{array}{ll}
 r^{-2l} (r^{\prime})^{2(l+1)} \ \mbox{\rm for $r^{\prime} \leq r$} \\
 r^{2(l+1)} (r^{\prime})^{-2l} \ \mbox{\rm for $r^{\prime} > r $} .
 \end{array} \right.
\end{equation}
Now for each $r \in I $:
\begin{equation} \label{comp}   
\int_{0}^{\infty} H(r, r^{\prime})dr{^\prime} = 
\frac{2(2l+1)}{(2l-1)(2l+3)} r^{3} |q(r)|^{2}
\end{equation}
and this expression is integrable over $I$ since $q$ is continuous and has a 
compact support. 
Hence the integrability of $H$ follows from Tonelli's Theorem. 
Collecting everything, we conclude that $B$ defines
a Hilbert-Schmidt operator on $L_{C}^2\left(I,r^4j\right)$. 

\end{document}